\documentclass[%
   preprint,
 showpacs,
 showkeys,
 preprintnumbers,
 amsmath,amssymb,
 aps,
  pra,
longbibliography,
 ]{revtex4-1}

\usepackage[dvipsnames]{xcolor}

\usepackage{mathptmx}

\usepackage{amssymb,amsthm,amsmath}

\usepackage{tikz}
\usetikzlibrary{calc}

\usepackage{graphicx}
\usepackage{url}

\usepackage{hyperref}
\hypersetup{
    colorlinks=true, 
    linktoc=all,     
    linkcolor=blue,  
    citecolor=blue,
    filecolor=red,
    urlcolor=blue
}

\begin{document}

\title{Comment on ``Quantum supremacy using a programmable superconducting processor''}

\author{Karl Svozil}
\email{svozil@tuwien.ac.at}
\homepage{http://tph.tuwien.ac.at/~svozil}

\affiliation{Institute for Theoretical Physics,
Vienna  University of Technology,
Wiedner Hauptstrasse 8-10/136,
1040 Vienna,  Austria}

\date{\today}

\begin{abstract}
Even if {\it Google AI}'s Sycamore processor is efficient for the particular task it has been designed for it fails to deliver universal computational capacity. Furthermore, even classical devices implementing transverse homoclinic orbits realize exponential speedups with respect to universal classical as well as quantum computations. Moreover, relative to the validity of quantum mechanics, there already exist quantum oracles which violate the Church-Turing thesis.
\end{abstract}

\keywords{quantum advantage, quantum parallelism, quantum random number generators, quantum oracles, classical chaotic systems, transverse homoclinic orbits}

\maketitle

Every system is a perfect and, oftentimes but not always, efficient {\em simulacrum} of itself.
Undoubtedly this seems to be the case for the particular task {\it Google AI}'s Sycamore processor~\cite{Arute2019}
has been designed for: sampling the output of a pseudo-random quantum circuit.
But does this justify claims of robust quantum ``supremacy'' (I prefer the term ``advantage'')?

The crucial question seems to be this:
what kind of computational tasks can be efficiently performed relative to
what kind of computational architecture?
The Church-Turing thesis states that
Turing machine type, as well as von Neumann-Zuse type architectures, can compute
what is informally acknowledged as an ``algorithm'' or ``computation''.
In this context two tasks appear as rather natural:
One task, inspired by computational complexity theory, is related to the assumption -- often referred to as Cook-Karp thesis --
that NP-complete problems are computationally intractable~\cite{garey}.
The other task is the effective simulation of universal computations in the sense of Church, Turing and Kleene~\cite{Smullyan1993-SMURTF}.

{\it Google AI}'s Sycamore processor yields no advantage for the first task and fails to support the second one.
One of the reasons for this is the very specialized computational capacities
which fail to render effective methods for solving NP-complete problems;
and even less so ``true'' universal computation.
In particular, the claim~\cite{Google-2019-qs} that Sycamore is a
{\em ``fully programmable 54-qubit processor''}
is unfounded if ``fully'' is understood in the
usual universal computable sense commonly used in theoretical computer science.

Let us retreat to the claim that Sycamore is efficient for the particular tasks it has been designed for.
And suppose for the moment the validity of the yet unproven conjecture that no classical computer
based on the aforementioned architectures can be as efficient
as Sycamore for those specific tasks.

Then it must be acknowledged that even classical ``processors'' exist which qualify for exponential advantages
relative to both the aforementioned architectures as well as to a wide variety of models of quantum compution:
classical chaotic systems, known since Poincare -- in particular, also systems as ``simple'' as  the (restricted) three-body problem --
realizing what is now called transverse homoclinic orbits~\cite{HOLMES1990137}
exhibit an exponential advantage over classical and quantum architectures due to their
sensitivity to variations of their initial state.
Such transverse homoclinic orbits cannot, in any non-construed way, universally compute.
And yet, to paraphrase claims made by {\it Google AI}'s Sycamore processor group,
{\em any physically realizable classical device realizing transverse homoclinic orbits (aka chaotic motion)
presents a polynomial-time computing machinery
for which no efficient method is known to exist for any universal computing machinery; both quantum and classical.}
This qualifies as a violation of the extended Church-Turing thesis formulated by Bernstein and Vazirani,
in a similar way as quoted for quantum sampling.

Indeed, if one postulates the validity of quantum mechanics, in particular,
complementarity and value indefiniteness (aka contextuality), then a genuine quantum advantage beyond the violation of the extended Church-Turing thesis
-- namely a violation of the original Church-Turing thesis --
has already been realized by numerous quantum random number generators.
Such devices are even commercially available.
Relative to the presumed irreducible randomness of quantum outcomes,
these quantum oracles for random numbers offer an absolute advantage
over any classical or quantum pseudorandom algorithm.



%

\end{document}